\documentclass[epsfig]{article}
\usepackage{graphicx}
\usepackage{amsmath}

\input epsf.sty
\input{tcilatex}
\begin{document}

\title{\textbf{Off-site interaction effect in the Extended Hubbard Model
with SCRPA method }}
\author{S. Harir$^{1\thanks{%
harirsaid@gmail.com}},$ M. Bennai$^{1,3\thanks{%
bennai\_idrissi@yahoo.fr}}$ and Y. Boughaleb$^{1,2,4}$ \\
$^{1}${\small \ Laboratoire de Physique de la Mati\`{e}re Condens\'{e}e,}\\
{\small Facult\'{e} des Sciences Ben M'Sik, Universit\'{e} Hassan
II-Mohammedia Casablanca,\ Morocco. }\\
$^{2}${\small \ LPMC, Facult\'{e} des Sciences d' El Jadida, Universit\'{e}
Chouaib Doukkali, Morocco.}\\
$^{3}${\small \ Groupement National de Physique des Hautes Energies,
LabUFR-PHE, Rabat, Morocco.}\\
$^{4}$ {\small Hassan II Academy of Sciences and Technology, Morocco}}
\date{}
\maketitle

\begin{abstract}
The Self Consistent Random Phase Approximation (SCRPA) and a Direct
Analytical (DA) method are proposed to solve the Extended Hubbard Model in
1D. We have considered an Extended Hubbard Model (EHM) including on-site and
off-site interactions for closed chains in one dimension with periodic
boundary conditions. The comparison of the SCRPA results with ones obtained
by a Direct Analytical approach shows that the SCRPA treats the problem of
these closed chains with a rigorous manner. The analysis of the
nearest-neighbour repulsion effect on the dynamics of our closed chains
shows that this repulsive interaction between the electrons of the
neighbouring atoms induces supplementary conductivity, since, the SCRPA
energy gap vanishes when these closed chains are governed by a strong
repulsive on-site interaction and intermediate nearest-neighbour repulsion.


\textbf{Pacs numbers:} 71.10, -W.75.10.Jm, 72.15.Nj
\end{abstract}

\section{Introduction}

\bigskip

The Hubbard Model is one of the most important models in the study of
strongly correlated systems \cite{Hub1,Hub2}. In spite of its simple
definition, it's believed to exhibit various interesting phenomena including
metal-insulator transition \cite{mott1,mott2,mott3,mott4,mott5},
antiferromagnetism \cite{Sanna,Nie,Antif3,antij}, and superconductivity \cite%
{Mc,Supra2}. In its description of interacting electrons on a lattice, the
Usual Hubbard Model exhibits the competition between the usual kinetic and
the on-site Coulomb interaction. In order, to establish a solution of the
self-consistent equations taking into account the correlations of type 2p-2h
(two particles- two holes), Jemai et al \cite{Jemai} proposed the Self
Consistent Random Phase Approximation (SCRPA). They used RPA method \cite%
{RPA,RPA2}, and introduced the RPA excitation operators for pair:
particle-hole (p-h), where the vacuum represents the ground state of the
system. These excitation operators are constructed from the linear
combinations of creation and annihilation operators of pair: p-h. The
quality of the SCRPA method to solve the Usual Hubbard Model has been
investigated in a previous work by Ref.\cite{Jemai}, in which, the authors
showed the remarkable agreement between the SCRPA and exact results.

To explain other physical phenomena observed in different areas of the solid
state physics like magnetic and transport properties, an Extended Hubbard
Model is proposed \cite{Calegari,Sol,Aich} taking into account the off-site
interaction, with supplementary parameter: $V$ (off-site interaction energy).

In this paper, we have analyzed the off-site interaction effect on the some
local properties of the 1D closed chains as: ground state energy, occupation
numbers and energy gap. The paper is organized as follows. In sec. 2, we
present the theory and describe, briefly, the SCRPA technique. In sec. 3, we
present the SCRPA results and compare them with ones obtained by a Direct
Analytical approach. Finally, we discuss the $V$ effect on the dynamics of
system.

\section{Theory}

\subsection{Model}

\bigskip The Extended Hubbard Hamiltonian is given by \cite{Calegari}:

\begin{equation}
H=\sum_{i\neq j,\sigma }t_{ij}c_{i,\sigma }^{^{\dagger }}c_{j,\sigma
}+U\sum_{i}n_{i,\uparrow }n_{i,\downarrow }+\frac{1}{2}\sum_{i\neq j,\sigma
,\sigma ^{\prime }}V_{ij}^{\sigma \sigma ^{\prime }}n_{i,\sigma }n_{j,\sigma
^{^{\prime }}}
\end{equation}%
The first term of the Eqs. $(1)$ represents the kinetic energy of electrons,
where each electron has a possibility of hopping between different lattice
sites. $c_{j,\sigma }$ is the annihilation operator of the electron at a
lattice site $j$ with spin index $\sigma $. $c_{i,\sigma }^{^{\dagger }}$ is
the creation operator of the electron at a lattice site $i$, so \ $t_{ij}$
is the hopping integral from the site $j$ to the site $i$. The second term
represents the on-site coulomb interaction with energy $U$, where $%
n_{i,\sigma }$ is the number operator of electrons at the site $i$ with spin 
$\sigma $ . $V_{ij}^{\sigma \sigma ^{\prime }}$ describes the effective
off-site coloumb interaction between the electrons in the lattice sites i
and j, with spin $\sigma $ and $\sigma ^{\prime }$respectively. The model
(1) cannot be solved in a general case. In this work we will limit ourselves
to a simplest case. We consider closed chains in one dimension with periodic
boundary conditions, where each closed chain is organized alternately of two
types of atoms noted "$A$" and "$B$" occupying the site $i$ and $i+1$,
respectively, and characterised by three parameters: $t$ (the first
nearest-neighbour hopping), $U$ (the on-site interaction energy) and $V$
(the off-site interaction energy). With $U\succ V\succ 0$, since the coulomb
interaction is always repulsive and decreases with the distance.

In the following of this paper, we suppose that $U\succ 2V$. In this case,
the $1D$ highly correlated metal have an anti-ferromagnetic order \cite{hir}%
. Thus, we have $\left\langle n_{i,\sigma }n_{i\pm 1,\sigma }\right\rangle
\simeq 0$. Under these conditions, the Hamiltonian of our physical system
is: 
\begin{equation}
H_{II}=-t\sum_{\sigma }(c_{i,\sigma }^{^{\dagger }}c_{i+1,\sigma
}+c_{i+1,\sigma }^{^{\dagger }}c_{i,\sigma })+U(n_{i,\uparrow
}n_{i,\downarrow }+n_{i+1,\uparrow }n_{i+1,\downarrow })+V\sum_{\sigma
}n_{i,\sigma }n_{i+1,-\sigma }
\end{equation}

In the first Brillouin zone $-\pi \leq k\prec \pi $, with the boundary
conditions, the possible wave numbers are: $k_{1}=0$ and $k_{2}=-\pi $.
Thus, the Hamiltonian of our system can be developed in Hartree-Fock ($HF$)
approximation as:

\begin{equation}
H_{HF}=E_{HF}+\sum_{\sigma }\left\{ \varepsilon _{1}\left( 1-a_{k_{1},\sigma
}^{^{\dagger }}a_{k_{1},\sigma }\right) +\varepsilon _{2}a_{k_{2},\sigma
}^{^{\dagger }}a_{k_{2},\sigma }\right\}
\end{equation}

$a_{k,\sigma }^{^{\dagger }}$ and $a_{k,\sigma }$ are, respectively, the
creation and annihilation operators of electrons with momenta $k$ and spin $%
\sigma $.

The Eqs. $(3)$ shows that, in the HF approximation, the system has two
possible excitation energies: $\epsilon _{1}=t-\frac{U+V}{2}$ ; $\epsilon
_{2}=t+\frac{U+V}{2}$ and the ground state energy is

\begin{equation*}
E_{HF}=\left\langle HF\right\vert H_{HF}\left\vert HF\right\rangle =-2\,t+%
\frac{U+V}{2}
\end{equation*}

Where $\left\vert HF\right\rangle =a_{k_{1},\uparrow }^{^{\dagger
}}a_{k_{1},\downarrow }^{\dagger }\left\vert vacuum\right\rangle $ is the $%
HF $ ground state of system. Whereas the $HF$ first exited state is taken
as: $\left\vert HF\right\rangle ^{\ast }=a_{k_{2},\uparrow }^{^{\dagger
}}a_{k_{2},\downarrow }^{\dagger }\left\vert vacuum\right\rangle $. Thus, we
can define the $HF$ quasiparticle operators by: $b_{1,\sigma
}=a_{k_{1},\sigma }^{^{\dagger }}$ and\ $b_{2,\sigma }=a_{k_{2},\sigma }$.
Therefore, we have $b_{k,\sigma }\left\vert HF\right\rangle =0$ for all $k$.

Finally , using the usual Fourier transformation for the operator $%
c_{i,\sigma }$, we can developed the Hamiltonian $H_{II}$ as function of the
operators $b_{1,\sigma }$ and $b_{2,\sigma }$.

\begin{equation}
H=H_{HF}+H_{k=0}+H_{k=-\pi }
\end{equation}

where 
\begin{equation*}
H_{k=0}=\frac{U+V}{2}\left( \tilde{n}_{k_{2},\uparrow }-\tilde{n}%
_{k_{1},\uparrow }\right) \left( \tilde{n}_{k_{2},\downarrow }-\tilde{n}%
_{k_{1},\downarrow }\right)
\end{equation*}%
\begin{equation*}
H_{k=-\pi }=-\frac{U-V}{2}\left( J_{\uparrow }^{-}+J_{\uparrow }^{^{\dagger
}})(J_{\downarrow }^{-}+J_{\downarrow }^{^{\dagger }}\right)
\end{equation*}

with

\begin{equation*}
J_{\sigma }^{-}=b_{1,\sigma }\,b_{2,\sigma },~J_{\sigma }^{^{\dagger
}}=\left( J_{\sigma }^{-}\right) ^{^{\dagger }},~\tilde{n}_{k_{i},\sigma
}=b_{i,\sigma }^{\dagger }\,b_{i,\sigma }
\end{equation*}

$H_{k=0}$ and $H_{k=-\pi }$ take into account, respectively, the correlation
between the number operators of the type: $\tilde{n}_{k_{i},\sigma }\tilde{n}%
_{k_{j},\sigma ^{\prime }}$ and between the magnetic moment operators of the
type: $J_{\sigma }^{^{\dagger }}.J_{\sigma ^{\prime }}^{-}$ .

\subsection{\protect\bigskip SCRPA approach}

The Random Phase Approximation (RPA) \cite{RPA}, was used to solve the Usual
Hubbard Model \cite{Jemai,Rabhi}. The RPA is an approach which treats
seriously the correlations of system, and attempt to minimise the system
energy. In order to apply the formalism of SCRPA to the Hubbard Model, it is
convenient to use the ph-RPA regrouping, which regroup the physical system
on pairs: particle-hole (p-h). We can then, define the RPA excitation
operator as:

\begin{equation}
Q_{v}^{^{\dagger }}=\sum_{p,h}(x_{ph}^{v}b_{p}^{^{\dagger }}b_{h}^{^{\dagger
}}-y_{ph}^{v}b_{h}b_{p})
\end{equation}%
Where $h$ (and $p$) are the momentum below (and above) the Fermi level. The
Eqs. $(5)$ shows that the excitation in the ph-RPA is done only by the
creation or (annihilation) of pair: particle-hole via the operator $%
b_{p}^{^{\dagger }}b_{h}^{^{\dagger }}$ ($b_{h}b_{p}$) with the amplitude $%
x_{ph}^{v}$ ($y_{ph}^{v}$). The corresponding excited state of this
excitation operator is $\left\vert v\right\rangle =Q_{v}^{^{\dagger
}}\left\vert RPA\right\rangle $, and the corresponding excitation energy is:

\begin{equation}
\varepsilon _{v}=\frac{\left\langle RPA\right\vert \left[ Q_{v},\left[
H,Q_{v}^{^{\dagger }}\right] \right] \left\vert RPA\right\rangle }{%
\left\langle RPA\right\vert \left[ Q_{v},Q_{v}^{^{\dagger }}\right]
\left\vert RPA\right\rangle }
\end{equation}%
Where $\left\vert RPA\right\rangle $ is the vacuum of this RPA excitation
operator: 
\begin{equation}
Q_{v}\left\vert RPA\right\rangle =0
\end{equation}%
We consider that this vacuum represents the ground state of the system. It's
clear, that the RPA approach will be more rigorous than the $HF$
approximation in the treatment of the correlation of system. Since, its
ground state is not defined explicitly like the $HF$ ground state, but, it
is defined only via the condition of the Eqs. $(7)$. Thus, the RPA ground
state can contain the correlations.

The minimization of $\varepsilon _{v}$ leads to usual RPA equations of type:

\begin{equation*}
\left( 
\begin{array}{cc}
A & B \\ 
-B^{\ast } & -A^{\ast }%
\end{array}%
\right) \left( 
\begin{array}{c}
x^{v} \\ 
y^{v}%
\end{array}%
\right) =\varepsilon _{v}\left( 
\begin{array}{c}
x^{v} \\ 
y^{v}%
\end{array}%
\right)
\end{equation*}

Where A and B are two square sub-matrix.

For the charge and spin longitudinal sector ($S=1$ , $m_{S}=0$)\cite{Jemai},
we consider only the RPA excitation operators which conserve the spin. Where
the excitation is done only by the creation or annihilation of the pair:
particle-hole with same spin.

\begin{equation}
Q_{v}^{^{\dagger }}=\frac{1}{\sqrt{\left\langle 1-M_{\uparrow }\right\rangle 
}}\left( x_{\uparrow }^{v}J_{\uparrow }^{^{\dagger }}-y_{\uparrow
}^{v}J_{\uparrow }^{-}\right) +\frac{1}{\sqrt{\left\langle 1-M_{\downarrow
}\right\rangle }}\left( x_{\downarrow }^{v}J_{\downarrow }^{^{\dagger
}}-y_{\downarrow }^{v}J_{\downarrow }^{-}\right)
\end{equation}

with 
\begin{equation*}
M_{\sigma }=\tilde{n}_{1,\sigma }+\tilde{n}_{2,\sigma }
\end{equation*}

In this sector, the sub-matrix A and B take the form:

$\ \ \ \ \ \ \ \ \ \ \ \ \ \ \ \ \ \ \ A=\left( 
\begin{array}{cc}
A_{\uparrow \uparrow } & A_{\uparrow \downarrow } \\ 
A_{\downarrow \uparrow } & A_{\downarrow \downarrow }%
\end{array}%
\right) $ \ \ \ \ \ \ \ \ \ \ \ \ \ \ \ and \ \ \ \ \ \ \ \ \ \ \ \ \ \ \ $%
B=\left( 
\begin{array}{cc}
B_{\uparrow \uparrow } & B_{\uparrow \downarrow } \\ 
B_{\downarrow \uparrow } & B_{\downarrow \downarrow }%
\end{array}%
\right) \ $

Where the RPA matrix elements are given by:

\begin{equation*}
A_{\sigma \sigma ^{\prime }}=\frac{1}{\sqrt{\left\langle 1-M_{\sigma
}\right\rangle \left\langle 1-M_{\sigma ^{\prime }}\right\rangle }}%
\left\langle \left[ J_{\sigma }^{-},\left[ H_{II},J_{\sigma ^{\prime
}}^{^{\dagger }}\right] \right] \right\rangle
\end{equation*}%
and

\begin{equation*}
B_{\sigma \sigma ^{\prime }}=\frac{1}{\sqrt{\left\langle 1-M_{\sigma
}\right\rangle \left\langle 1-M_{\sigma ^{\prime }}\right\rangle }}%
\left\langle \left[ J_{\sigma }^{-},\left[ H_{II},J_{\sigma ^{\prime }}^{-}%
\right] \right] \right\rangle
\end{equation*}

With the relations of the orthonormality conditions of the set $\left\{
Q_{v};Q_{v^{\prime }}^{^{\dagger }}\right\} $, we have expressed the
elements of $A$ and $B$ by the RPA-amplitudes, and therefore we have
constructed a closed system of non linear coupled equations, which we have
solved numerically by iteration leading to the SCRPA solutions. For more
details, see Ref. \cite{applied}. In the present paper, we will compare
these SCRPA solutions with ones obtained by a Direct Analytical approach for
two local properties of our closed chain: the SCRPA ground state energy: $%
E_{_{GS}}^{^{SCRPA}}=\left\langle RPA\right\vert H_{II}\left\vert
RPA\right\rangle $ and the occupation number of states with spin $\sigma $
and momenta $k$ below the Fermi level : $n_{k\prec k_{F},\sigma
}^{^{SCRPA}}=1-\left\langle RPA\right\vert J_{\sigma }^{^{\dagger
}}J_{\sigma }^{-}\left\vert RPA\right\rangle $.

\section{Results and discussion}

In this section, we present our SCRPA results and compare them with ones
obtained by a Direct Analytical (DA) method.

\subsection{DA method}

To take into account the correlation of the first order for the ground
state, we suppose that the Direct Analytical ground state wave function
contains the correlation of type $(2p-2h)$, therefore the ground state is: 
\begin{equation*}
\left\vert 0\right\rangle =\left( c_{0}^{v}+c_{\uparrow \downarrow
}^{v}.J_{\uparrow }^{^{\dagger }}J_{\downarrow }^{^{\dagger }}\right)
\left\vert HF\right\rangle
\end{equation*}

The Hamiltonian of system in the subspace \{ $\left\vert HF\right\rangle $ $%
\ $;$\ J_{\uparrow }^{^{\dagger }}J_{\downarrow }^{^{\dagger }}.\left\vert
HF\right\rangle $ \ \} is:

\begin{equation}
H=\left( 
\begin{array}{cc}
-2t+\alpha & -\beta /2 \\ 
-\beta /2 & 2t+\alpha%
\end{array}%
\right)
\end{equation}

Where 
\begin{equation*}
\alpha =U+V\text{ \ \ \ and \ \ \ \ \ \ \ }\beta =U-V
\end{equation*}

This Hamiltonian has two roots:

\begin{equation*}
E_{0}=\alpha -\frac{\sqrt{\beta ^{2}+16.t^{2}}}{2}
\end{equation*}

and

\begin{equation*}
\text{\ }E_{1}=\alpha +\frac{\sqrt{\beta ^{2}+16.t^{2}}}{2}
\end{equation*}

We take the eigenvector corresponding to $E_{0}$ as the exact ground state,
we have then:

\begin{equation}
\left\vert 0\right\rangle =\left( \cos (\phi )+\sin (\phi )J_{\uparrow
}^{^{\dagger }}J_{\downarrow }^{^{\dagger }}\right) \left\vert
HF\right\rangle
\end{equation}

with%
\begin{equation*}
\phi =\arctan (\frac{\text{\ }\beta }{4t+\sqrt{\text{\ }\beta ^{2}+16t^{2}}})
\end{equation*}

Finally, the Direct Analytical ground state energy and the occupation number
of states with spin $\sigma $ and momenta $k$ below the Fermi level are
calculated, respectively, via:

\begin{equation*}
E_{_{GS}}^{^{DA}}=\left\langle 0\right\vert H\left\vert 0\right\rangle \text{
\ \ \ \ \ \ \ and \ \ \ \ \ }n_{k\prec k_{F},\sigma }^{^{DA}}=1-\left\langle
0\right\vert J_{\sigma }^{^{\dagger }}J_{\sigma }^{-}\left\vert
0\right\rangle
\end{equation*}

Using the Eqs. $(9)$ and $(10)$, we can find:

\begin{equation}
E_{_{GS}}^{^{DA}}=-2t\cos (2\phi )+\frac{U}{2}(1-\sin (2\phi ))+\frac{V}{2}%
(1+\sin (2\phi ))
\end{equation}

\begin{equation}
n_{k\prec k_{F},\sigma }^{^{DA}}=\cos ^{2}(\phi )
\end{equation}

Where, the parameter $\phi $ is given by: 
\begin{equation*}
\ \phi =\arctan (\frac{U-V}{4t+\sqrt{(U-V)^{2}+16t^{2}}})
\end{equation*}

\subsection{SCRPA and DA results}

In Fig.1 (Fig.2), we have plotted the variation of $E_{_{GS}}$ and $%
n_{k\prec k_{F},\sigma }$ of SCRPA and DA methods as function of $U/t$ for
different values of $V/t$ .\FRAME{ftbpFU}{5.6654in}{4.1425in}{0pt}{\Qcb{%
Ground state energy of SCRPA and DA methods as function of on-site
interaction energy $U/t$ for different values of $V/t$.}}{}{jfpf4002.wmf}{%
\special{language "Scientific Word";type "GRAPHIC";maintain-aspect-ratio
TRUE;display "USEDEF";valid_file "F";width 5.6654in;height 4.1425in;depth
0pt;original-width 5.6074in;original-height 4.0932in;cropleft "0";croptop
"1";cropright "1";cropbottom "0";filename '../JFPF4002.wmf';file-properties
"XNPEU";}}

\FRAME{ftbpFU}{5.7795in}{3.8795in}{0pt}{\Qcb{Occupation number of states
with spin $\protect\sigma $ and momenta $k$ below the Fermi level of SCRPA
and DA methods as function of on-site interaction energy $U/t$ for different
values of $V/t$. \ }}{}{jjz6fv00.wmf}{\special{language "Scientific
Word";type "GRAPHIC";maintain-aspect-ratio TRUE;display "USEDEF";valid_file
"F";width 5.7795in;height 3.8795in;depth 0pt;original-width
5.7207in;original-height 3.8311in;cropleft "0";croptop "1";cropright
"1";cropbottom "0";filename '../JJZ6FV00.wmf';file-properties "XNPEU";}}

\bigskip The Fig.1 and Fig.2 show that the agreement between the SCRPA and
DA methods is very good for any value of $U/t$ and $V/t$. It is clear that
the SCRPA method loses partially its precision for the high values of $U/t$,
but as same on an important interval $\left[ 0\text{ };\text{ }6t\right] $
the method keeps all its precision. We can conclude, thus, that the SCRPA
method is efficient to treat seriously the off-site interaction of type $%
n_{i,\sigma }n_{i+1,-\sigma }$. The curves show, also, that for $U\prec V$,
we have no SCRPA solution. The program associated to SCRPA method begins to
turn and give the solutions for $U\succeq V$; whereas the DA method gives
solutions even for $U\prec V$. Thus, we can conclude that the SCRPA method
is more realist than the DA method; knowing that for a real physical system,
the on-site interaction is always stronger than the off-site interaction ($%
U\succ V$).

In Fig. 3, we have plotted the variation of $E_{_{GS}}$ of SCRPA and DA
methods as function of $V/t$ at $U=2t$.

The Fig. 3 shows that the SCRPA and DA results are practically equal for any
value of V/t. Noting that the program associated to the SCRPA stops to give
the solutions, when $V$ becomes the order of $U$ ($V\simeq U$). This figure
shows, also, that the ground state energy of our closed chains has,
practically, a linear V dependence.

\FRAME{ftbpFU}{5.9317in}{3.8917in}{0pt}{\Qcb{Ground state energy of SCRPA
and DA methods as function of off-site interaction energy V=t at U = 2t.}}{}{%
jjz7ta01.wmf}{\special{language "Scientific Word";type
"GRAPHIC";maintain-aspect-ratio TRUE;display "USEDEF";valid_file "F";width
5.9317in;height 3.8917in;depth 0pt;original-width 5.8721in;original-height
3.8432in;cropleft "0";croptop "1";cropright "1";cropbottom "0";filename
'../JJZ7TA01.wmf';file-properties "XNPEU";}}

In summary, we have compared the SCRPA results with ones obtained by a DA
method, we have found that SCRPA method solves the Extended Hubbard Model
for closed chains in 1D exactly for any value of \ $U$ and $V$.

\subsection{\protect\bigskip Effect of $V$}

In order to analyse the effect of $V$ on the dynamics of our closed chains,
we define the energy gap $\Delta _{E}$\ as the difference between the first
excited state energy $\epsilon _{1}$\ \cite{applied} and the ground state
energy $E_{SCRPA}$:

\begin{equation*}
\Delta _{E}=\epsilon _{1}-E_{SCRPA}
\end{equation*}

In Fig.4, we plot the variation of the energy gap $\Delta _{E}$ as function
of the repulsive on-site interaction energy $U$ for different values of the
off-site interaction energy $V$, with $2V\prec U$.

\FRAME{ftbpFU}{5.668in}{3.8562in}{0pt}{\Qcb{Energy gap as function of
on-site interaction energy $U/t$ for different values of $V/t$}}{}{%
jjx72600.wmf}{\special{language "Scientific Word";type
"GRAPHIC";maintain-aspect-ratio TRUE;display "USEDEF";valid_file "F";width
5.668in;height 3.8562in;depth 0pt;original-width 5.61in;original-height
3.8069in;cropleft "0";croptop "1";cropright "1";cropbottom "0";filename
'../JJX72600.wmf';file-properties "XNPEU";}}

First, we disregarded the off-site interaction: $V=0$ (Usual Hubbard Model
case ), the corresponding curve shows that the gap energy $\Delta _{E}$
decreases when $U$ increases. We deduce that the repulsive on-site
interaction $(U\succ 0)$ increase the conductivity of system, since the
repulsion between the two electrons of one site encourages every electron to
jump to the neighbouring site. Then, we have introduced the off-site
interaction $(V\neq 0)$, and we have plotted the variation of the energy gap 
$\Delta _{E}$ as function of the repulsive on-site interaction energy $U$
for different values of the\ interaction off-site energy: $V/t$. The curves
show that the effect of $V$ \ on the energy gap $\Delta _{E}$ is very weak
when the on-site interaction is weak. For instance, at $U\simeq 2t$ the
effect of $V$ on $\Delta _{E}$ is the order of $18\%$ for an intermediate
off-site interaction $\left( V/t=1\right) $. Whereas, when the on-site
interaction energy increases, the effect of this off-site interaction will
be more remarkable. For instance, at $U\simeq 6t$ (strong on-site
interaction ) the effect of this intermediate off-site interaction $\left(
V/t=1\right) $ on $\Delta _{E}$ will be the order of\ $50\%$. The curves
show also that when $U$ increases, the gap energy $\Delta _{E}$ decreases
and tend to a constant value $\Delta _{0}$ at critical on-site interaction
energy $U_{c}$. This limit value $\Delta _{0}$ depends on off-site
interaction energy \ $V$, noticing that $\Delta _{0}$ vanishes for the
intermediate repulsive off-site interaction $\left( V/t=1\right) $. We
deduce that the repulsive off-site interaction increases the conductivity of
system, especially, in the case of an intermediate repulsive off-site
interaction, the gap between the first excited state and the ground state
energy vanishes, and therefore each electron can jump between different
sites without any loss of energy.

\section{Conclusion}

In this paper, we have used the SCRPA and DA methods to solve the 1D
Extended Hubbard Model in order to study some local properties of closed
chains in one dimension with periodic boundary conditions as: ground state
energy, energy gap and occupation numbers. First, We have compared the SCRPA
results with ones obtained by the Direct Analytical method; we have shown
that the SCRPA method solves the Extended Hubbard Model for finite number of
sites exactly, for any value of $U$ (on-site interaction energy) and $V$
(off-site interaction energy). Then, we have analysed the $V$ effect on the
dynamics of system. We have shown that the energy gap decreases with $V$.
Particularly, we have found that with an intermediate repulsive off-site
interaction ($V/t=1$) this gap between the first excited state energy and
the ground state energy vanishes, and therefore the electrons can jump
between different sites without any loss of energy.

\end{document}